\documentclass[runningheads]{llncs}
\usepackage{amsmath}
\usepackage{mathtools}
\usepackage{graphicx}
\usepackage{graphics}
\usepackage{acro}

\usepackage[colorinlistoftodos]{todonotes}
\usepackage[colorlinks=true, allcolors=blue]{hyperref}
\usepackage{caption}
\DeclareCaptionFont{mysize}{\fontsize{9}{9}\selectfont}
\usepackage[title]{appendix}
\usepackage{amsfonts}
\usepackage{multirow}
\usepackage{booktabs}
\usepackage[ruled,vlined,algo2e]{algorithm2e}

\DeclareAcronym{CT}{
short=CT,
long=computed tomography
}
\DeclareAcronym{PACS}{
short=PACS,
long=picture archiving and communication system
}
\DeclareAcronym{SE}{
short=SE,
long=squeeze and excitation
}
\DeclareAcronym{ACE}{
short=ACE,
long=aggregated cross entropy
}

\DeclareAcronym{CE}{
short=CE,
long= cross entropy
}
\DeclareAcronym{NC}{
short=NC,
long=non-contrast
}
\DeclareAcronym{D}{
short=D,
long=delay
}
\DeclareAcronym{A}{
short=A,
long=arterial
}
\DeclareAcronym{V}{
short=V,
long=venous
}
\DeclareAcronym{O}{
short=O,
long=other
}
\DeclareAcronym{NLP}{
short=NLP,
long=natural language processing
}

\DeclareAcronym{SOI}{
short=SOI,
long=scan of interest,
long-plural-form=scans of interest
}

\def\@fnsymbol#1{\ensuremath{\ifcase#1\or *\or \dagger\or \ddagger\or
   \mathsection\or \mathparagraph\or \|\or **\or \dagger\dagger
   \or \ddagger\ddagger \else\@ctrerr\fi}}
\newcommand{\ssymbol}[1]{^{\@fnsymbol{#1}}}

\newcommand{\eg}{\emph{e.g.},}
\newcommand{\ie}{\emph{i.e.},}
\newcommand{\etal}{\emph{et al.}}
\newcommand{\Fig}{Fig.}
\newcommand{\Tbl}{Tbl.}
\DeclareMathOperator{\logsumexp}{logsumexp}
\begin{document}

\title{CT Data Curation for Liver Patients: Phase Recognition in Dynamic Contrast-Enhanced CT}
 
\author{Bo Zhou\inst{1,2} \and Adam P. Harrison\inst{2} \and Jiawen Yao\inst{2} \and Chi-Tung Cheng\inst{4} \and Jing Xiao\inst{3} \and Chien-Hung Liao\inst{4} \and Le Lu\inst{2}}
\institute{Biomedical Engineering, Yale University, New Haven, CT, USA \and PAII Inc., Bethesda, MD, USA \and PingAn Technology, Shenzhen, China \and Chang Gung Memorial Hospital, Linkou, Taiwan, ROC}

\authorrunning{B. Zhou, A. P. Harrison, etc} 
\titlerunning{Data Curation for Patients: Phase Recognition} 

\maketitle              

\begin{abstract}
As the demand for more descriptive machine learning models grows within medical imaging, bottlenecks due to data paucity will exacerbate. Thus, collecting enough large-scale data will require automated tools to harvest data/label pairs from messy and real-world datasets, such as hospital \acp{PACS}. This is the focus of our work, where we present a principled data curation tool to extract multi-phase \ac{CT} liver studies and identify each scan's phase from a real-world and heterogenous hospital \ac{PACS} dataset. Emulating a typical deployment scenario, we first obtain a set of noisy labels from our institutional partners that are text mined using simple rules from DICOM tags. We train a deep learning system, using a customized and streamlined 3D \ac{SE} architecture, to identify non-contrast, arterial, venous, and delay phase dynamic \ac{CT} liver scans, filtering out anything else, including other types of liver contrast studies. To exploit as much training data as possible, we also introduce an aggregated cross entropy loss that can learn from scans only identified as ``contrast''. Extensive experiments on a dataset of 43K scans of 7680 patient imaging studies demonstrate that our 3DSE architecture, armed with our aggregated loss, can achieve a mean F1 of $0.977$ and can correctly harvest up to $92.7\%$ of studies, which significantly outperforms the text-mined and standard-loss approach, and also outperforms other, and more complex, model architectures. 
\end{abstract}
\acresetall

\vspace{-0.6cm}
\section{Introduction}
Over the last decade, deep learning techniques have seen success in automatically interpreting biomedical and diagnostic imaging data~\cite{Litjens_2018,zhou2018generation}. However, robust performance often requires training from large-scale data. Unlike computer vision datasets, which can rely on crowd-sourcing~\cite{deng2009imagenet}, the collection of large-scale medical imaging datasets must typically involve physician labor. Thus, there exists a tension between modeling power and data requirements that only promises to increase~\cite{Kohli_2017}. An enticing prospect is mining physician expertise by collecting retrospective data from \acp{PACS}, but the current generation of \acp{PACS} do not properly address the curation of large-scale data for machine learning. In \acp{PACS}, DICOM tags regarding scan descriptions are typically hand inputted, non-standardized, and often incomplete, which leads to the need for extensive data curation \cite{harvey2019standardised}. These limitations frequently produce high mislabeling rates, \eg{} the $15\%$ rate reported by Gueld \etal{}, meaning that simply selecting the \acp{SOI} from a large set of studies can be prohibitively laborious. This has spurred efforts to automatically text mine image/label pairs from \acp{PACS}~\cite{yan2018deeplesion,zhou2019progressively,Irvin_2019}, but these efforts rely on complicated and customized \ac{NLP} technology to extract labels. Apart from the barriers put forth by this complexity, these solutions address contexts where it is possible to extract the information of interest from accompanying text. This is not always possible, as \ac{NLP} parsers~\cite{Peng_2018,Irvin_2019} cannot always straightforwardly correct errors in the original reports or fill in missing information. As such, collecting large-scale data will also require developing automated, but robust, tools that go beyond mining from DICOM tags and/or reports. 

\begin{figure}[!htb]
\centering
\includegraphics[width=0.92\textwidth]{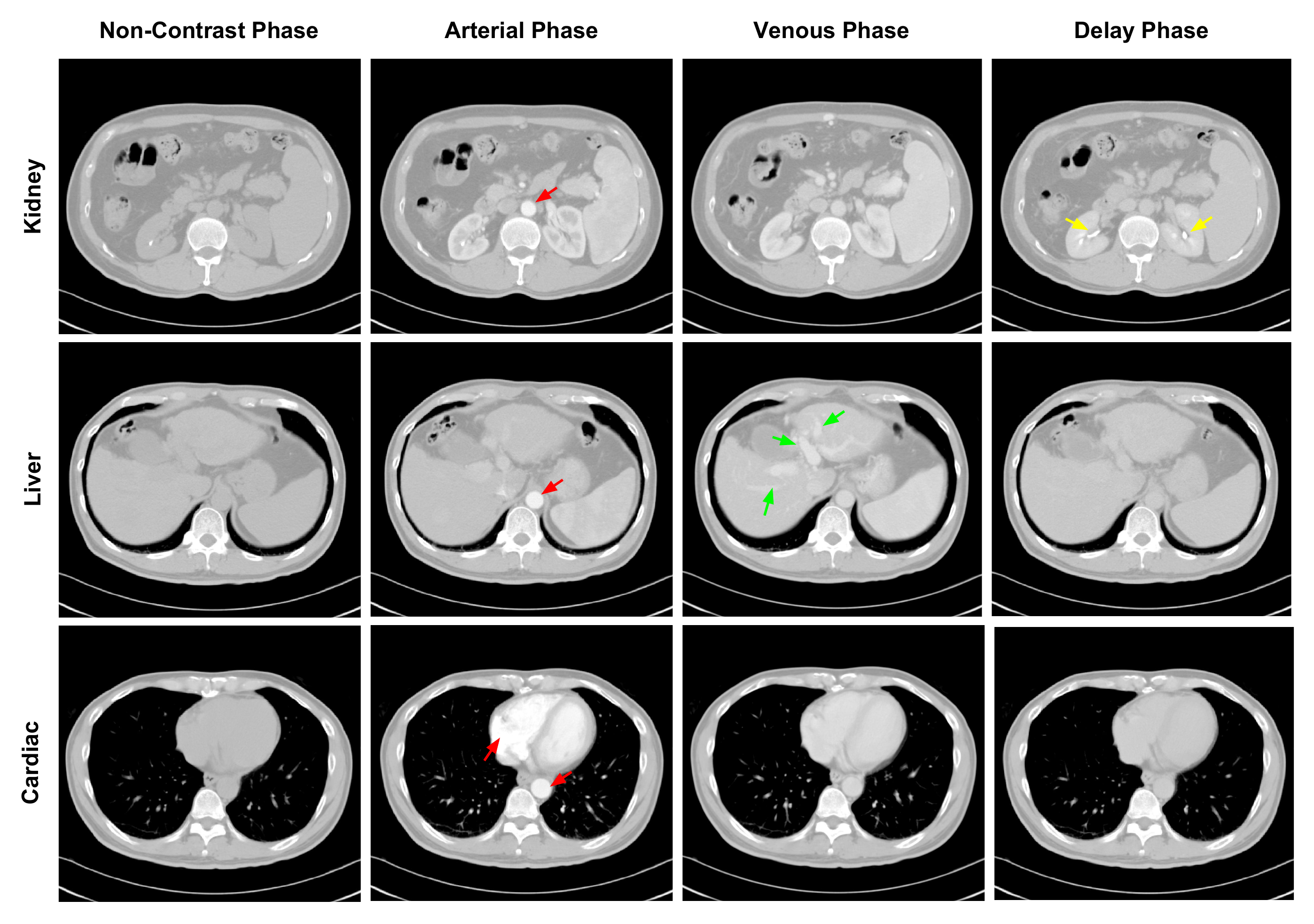}
\caption{\Acf{NC}, \acf{A}, \acf{V}, and \acf{D} phases are the \acp{SOI} in dynamic CT. Radiologists use contrast information in several organs to determine the phase, such as contrast in the heart/aorta (red arrows), portal veins (green arrows), and kidneys (yellow arrows).}
\label{fig:phase_ill}
\end{figure}

This is the topic of our work, where we articulate a robust approach to large-scale data curation based on visual information. In our case, we focus on a hospital \ac{PACS} dataset we collected that consists of $43\,010$ \ac{CT} scans of $7\,680$ imaging studies from $4\,666$ unique patients with liver lesions, along with pathological diagnoses. Its makeup is highly heterogeneous, comprising studies of multiple organs, protocols, and reconstruction types. Very simple and accessible text matching rules applied to the DICOM tags can accurately extract scan descriptions; however omissions and errors in the text mean these labels are noisy and unreliable. Without loss of generality, we focus on extracting a large-scale and well curated dataset of dynamic liver \ac{CT} studies from our \ac{PACS} data. Dynamic \ac{CT} is the most common protocol to categorize and assess liver lesions~\cite{burrowes2017contrast}, and we expect a large-scale dataset to prove highly valuable for the development of computer-aided diagnosis systems, \textit{provided it is well curated}. Thus, the goal is to use the noisy labels to train a visual recognition system that can much more robustly identify dynamic liver \ac{CT} studies, extract the corresponding axial-reconstructed scans, and identify the phase of each as being \ac{NC}, \ac{A}, \ac{V}, or \ac{D}. Fig.~\ref{fig:phase_ill} shows examples of each phase and discriminating features of each.

Unlike prior work, we focus on extracting multi-phase volumetric \acp{SOI} of a certain type, rather than on extracting disease tags or labels. This places a high expectation on performance, \ie{} F1 scores of $0.95$, or higher. To tackle this problem, we develop a principled phase recognition system whose contributions are threefold. First, we collect the aforementioned large-scale dataset from a hospital \ac{PACS}, that includes more than $43\,010$ scans. Second, we introduce a customized phase-recognition deep-learning model, comprised of a streamlined version of C3D~\cite{tran2015learning} with \ac{SE} layers. We show that this simple, yet effective model, can outperform much more complicated models. Third, we address a common issue facing data curation systems, where many text mined labels are too general. In our case, these are labels that indicate only ``contrast'' rather than the more specific \ac{NC}, \ac{A}, \ac{V}, or \ac{D} \acp{SOI}. So that we can still use these images for training, along with their weak supervisory signals, we design an \ac{ACE} loss that incorporates the hierarchical relationship within annotations. Our experimental results demonstrate that our 3DSE model, in combination with our \ac{ACE} loss, can achieve significantly better phase recognition performance than the text-mined method and other deep-learning based approaches. To the best of our knowledge, this is the first work investigating visual-information based data curation methods in \ac{PACS}, and we expect that our data curation system would also prove a useful curation approach in domains other than liver dynamic \ac{CT}.

\vspace{-0.2cm}
\section{Methods}
\vspace{-0.1cm}
\subsection{Dataset}
Our goal is to reliably curate as large as possible a dataset of liver dynamic \ac{CT} scans, with minimal labor. To do this, we first extracted a dataset of \ac{CT} studies from the \ac{PACS} of \emph{Anomymized}, corresponding to patients who had pathological diagnoses of liver lesions, with the hope that such a dataset would be of great interest for later downstream analysis. This resulted in $7\,680$ studies of $4\,666$ patients. For each study, the number of scans range from $4$ to $30$ and there are one to three studies per patient. The resulting dataset is highly heterogenous, containing several types of reconstructions, projections, anatomical regions, and contrast protocols that we not interested in, \eg{} computed tomography arterial portography. Studies containing dynamic \ac{CT} scans may have anywhere from one or all of \ac{NC}, \ac{A}, \ac{V}, and \ac{D} contrast phase \acp{SOI}. Our aim is to identify and extract the axial-reconstructed versions of these scans from each study, should they exist. As such, this task exemplifies many of the general demands and challenges of data curation across medical domains. 

With the dataset collected, we next applied a set of simple text matching rules to the DICOM tags to noisily label each scan as being either \ac{NC}, \ac{A}, \ac{V}, \ac{D} or \ac{O}. The full set of rules are tabulated in our supplemental materials. The text-matching rules are more than sufficient to reliably extract labels \emph{based on text alone}, due to the extremely simple structure and vocabulary of DICOM tags. However, because the source DICOM tags are themselves error-prone and unreliable~\cite{Gueld_2017}, these labels suffer from inaccuracies, which we demonstrate later in our results. Finally, we filter out any scans that have less than $10$ slices, with a spatial resolution coarser than $5\mathrm{mm}$, or were taken after or during a biopsy or transplant procedure. As a result, we found 1728, 1703, 1504 and 1736 \ac{A}, \ac{V}, \ac{D} and \ac{NC} scans, respectively, with 326 scans labeled as ‘contrast’. We then manually annotated a validation set and a test set, comprising $801$ and $1262$ scans;  $150$ and $231$ studies; and $101$ and  $196$ patients, respectively. This left a training set of $29\,891$ scans from $5\,164$ studies of $3\,267$ patients with noisy text-mined annotations. 

\vspace{-0.1cm}
\subsection{3DSE Network}
As \Fig~\ref{fig:phase_ill} illustrates, visual cues indicating the phase can be located in different anatomical areas. Given this, we opt for a 3D classification network. State of the art 3D classification networks, such as 3D-Resnet~\cite{hara2017learning} and C3D~\cite{tran2015learning}, are often quite large, adding to the training time and increasing overfitting tendencies. 

\begin{figure}
\centering
\includegraphics[width=0.95\textwidth]{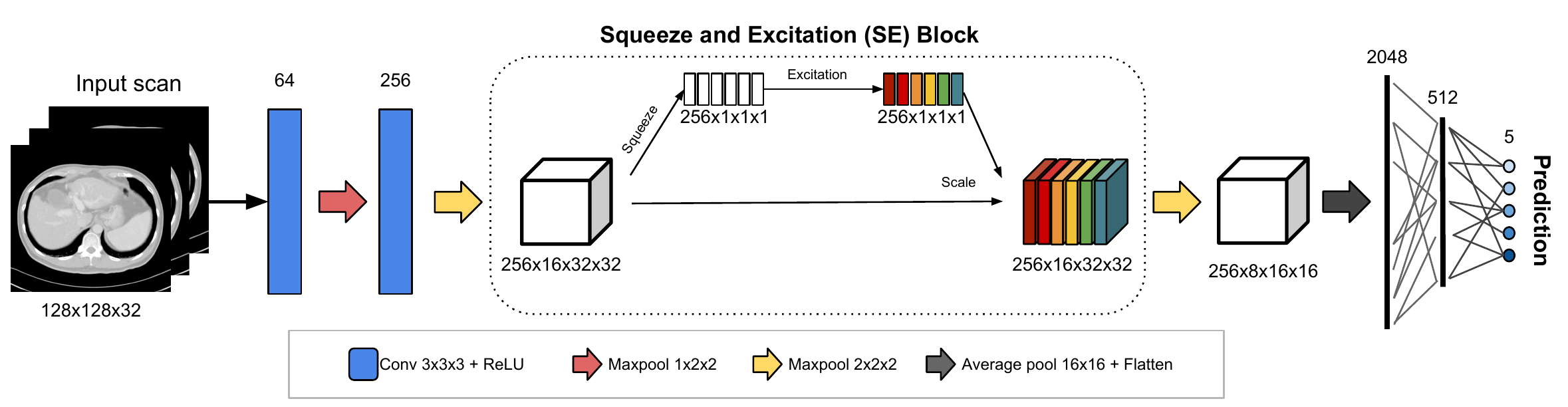}
\caption{Our 3DSE network is designed to have a relatively small amount of parameters and consists of three parts, including two 3D convolution layers, one \acs{SE} layer, and two fully connected layers.}
\label{fig:3dsenet}
\end{figure}

Instead, we use a streamlined but effective architecture we call 3DSE, which is illustrated in \Fig~\ref{fig:3dsenet}. To begin, we first downsample all volumes to $128\times128\times32$. From these, image features are extracted using two convolutional layers, each followed by a rectified linear unit and max pooling layers. With such a streamlined feature extracter, activation maps are highly local~\cite{hu2018squeeze}. Thus, we add \acf{SE}~\cite{hu2018squeeze} layers. These scale each feature channel with multiplicative factors computed using global pooling, providing an efficient means to increase descriptive capacity and inject global information. Subsequent pooling layers and a two fully connected layers provide the five output phase predictions. The total parameter size $19.22$ MB which is significantly smaller than 3D-Resnet~\cite{hara2017learning} and C3D~\cite{tran2015learning}.

\vspace{-0.1cm}
\subsection{Aggregated Cross Entropy}
Frequently, text-mined labels are only able to provide a more general label of ``contrast'' for a scan, indicating that it could be any of \ac{A}, \ac{V}, or \ac{D} \acp{SOI}. Since our goal is to determine the exact phase, the easiest way to handle such scans is to simply remove them from training, at the cost of using less data. Yet, such weakly supervised data still provides useful information, which should ideally be exploited to use as much training data as possible. To do this, we formulate a simple \acf{ACE} loss that can execute a \ac{CE} loss, but these weakly supervised instances. We formulate the probability of ``contrast'' as equalling the sum of the probabilities of all contrast phases:
\begin{align}
p_{\mathrm{C}} &= p_{\mathrm{A}}+p_{\mathrm{V}}+p_{\mathrm{D}} \textrm{,} \\
&= \dfrac{\exp(\mathbf{w}_{\mathrm{A}}) + \exp(\mathbf{w}_{\mathrm{V}}) + \exp(\mathbf{w}_{\mathrm{D}})}{\sum_{i}\exp(\mathbf{w}_{i})} \textrm{,} \label{eqn:p_c_2}
\end{align}
where \eqref{eqn:p_c_2} assumes a pseudo-probability calculated using softmax, $\mathbf{w}_{(.)}$ denotes the logit outputs, and $i$ indexes all five outputs. 

The $p_{C}$ can be naively used in a \ac{CE} loss, but that would preclude using a numerically stable ``softmax with \ac{CE}'' formulation. Instead, for scans that can only be labelled as ``contrast'', the \ac{CE} loss can be written as:
\begin{align}
\ell_{CE}&=-y_{\mathrm{NC}}\log(p_{\mathrm{NC}}) - y_{\mathrm{O}}\log(p_{\mathrm{O}}) - y_{\mathrm{C}}\log(p_{\mathrm{C}}) \textrm{,} \\
&=-\log\left(\dfrac{\exp(\mathbf{w}_{\mathrm{A}}) + \exp(\mathbf{w}_{\mathrm{V}}) + \exp(\mathbf{w}_{\mathrm{D}})}{\sum_{i}\exp(\mathbf{w}_{i})}\right) \textrm{,} \label{eqn:ace_2}\\
&=\logsumexp(\{\mathbf{w}_{i}\}) - \logsumexp(\{\mathbf{w}_{\mathrm{A}}, \mathbf{w}_{\mathrm{V}}, \mathbf{w}_{\mathrm{D}\}}) \textrm{,} \label{eqn:ace_3}
\end{align}
where $y_{(.)}$ denotes the ground truth. The elimination of all terms but the contrast term in \eqref{eqn:ace_2}, follows from $y_{\mathrm{C}}$ equalling one, with all other  $y_{(.)}$ values equalling zero. The $\logsumexp$ function enjoys numerically stable forward- and backward-pass implementations. Thus, when presented with a ``contrast'' scan, our model uses \eqref{eqn:ace_3} for the loss, providing a simple and numerically stable means to exploit all available data to train our desired, but more fine-grained, outputs. 

\vspace{-0.2cm}
\section{Results}

We tested our 3DSE network, with and without the \ac{ACE} loss, on our dataset, and compared it to both the noisy text-mined labels and also 3D-Resnet-101~\cite{hara2017learning} and C3D~\cite{tran2015learning}. For all models we perform a sweep of learning rates and report results corresponding to the best setting and stopping point based on the validation set. 

Focusing first on scan-level comparisons, \Tbl~\ref{tab:scan_metrics} presents F1 scores across the different phase types. 
\begin{table}
\centering
\caption{Quantitative comparison of scan-level performance. Best results are marked in \textcolor{blue}{blue}. For the 3DSE + \acs{ACE} F1 phase-level scores, we use {\@fnsymbol{1}} and {\@fnsymbol{2}} to indicate if differences were statistically significant ($\alpha <0.05$) compared to the text-mining and 3DSE model, respectively. Significance was calculated using randomized tests~\cite{Yeh_2000} and adjusted using the multiple comparison correction of Holm-Bonferroni~\cite{Holm_1979}.}
\label{tab:scan_metrics}
    \begin{tabular}{|c|c|c|c||c|c|c||c|c|c|}
        \hline
           & \multicolumn{3}{c||}{Text Mining} & \multicolumn{3}{c||}{3DSE} & \multicolumn{3}{c|}{3DSE + \acs{ACE}} \\
        \hline
           & Precision  & Recall & F1 Score & Precision  & Recall & F1 Score & Precision  & Recall & F1 Score \\ [0.025cm]
        \hline
        NC & 0.977 & 0.895 & 0.934 & 0.965 & 0.965 & 0.964 & 0.993 & 0.986 & \textcolor{blue}{0.988}$\ssymbol{1}$$\ssymbol{2}$ \\ [0.025cm]
        \hline
        A  & 0.966 & 0.983 & 0.974 & 0.974 & 0.966 & 0.970 & 0.991 & 0.991 & \textcolor{blue}{0.992} \\ [0.025cm]
        \hline
        V  & 0.975 & 0.782 & 0.868 & 0.965 & 0.946 & 0.956 & 0.930 & 0.993 & \textcolor{blue}{0.963}$\ssymbol{1}$ \\ [0.025cm]
        \hline
        D  & 0.964 & 0.956 & 0.960 & 0.964 & 0.956 & \textcolor{blue}{0.960} & 0.972 & 0.930 & 0.951 \\ [0.025cm]
        \hline
        O  & 0.926 & 0.986 & 0.955 & 0.981 & 0.989 & 0.985 & 0.997 & 0.990 & \textcolor{blue}{0.993}$\ssymbol{1}$$\ssymbol{2}$ \\ [0.025cm]
        \hline
        mean  & 0.962 & 0.920 & 0.938 & 0.970 & 0.964 & 0.967 & 0.977 & 0.978 & \textcolor{blue}{\textbf{0.977}} \\ [0.025cm]
        \hline
    \end{tabular} \vspace{-2.6mm}
\end{table}
As can be observed from the text-mined results, many scans are misclassified as \ac{O} and many \ac{D} scans are missed, demonstrating the shortfalls of relying on labels based on DICOM tags. In contrast, the vision-based 3DSE significantly reduces classification errors, improving the mean F1 score from $0.938$ (via text mining) to $0.967$. In particular, \ac{V}'s F1 score is improved from $0.868$ to $0.956$. Performance is increased even further when we use the \ac{ACE} loss to include the ``contrast'' scans in training, boosting the mean F1 score to $0.977$. While tests show a degradation of performance for the \ac{D} phase, these differences do not meet statistical significance, unlike the statistically significant improvements seen in the \ac{NC}, \ac{V}, and \ac{O} phases. Thus, these results validate the use of our \ac{ACE} formulation to exploit as much training data as possible. 
\begin{table}[t]
\centering
\caption{Across-model quantitative evaluation using the F1 score. Best and second-best results are marked in \textcolor{blue}{blue} and \textcolor{red}{red}, respectively.}
\label{tab:cnn_compare}
    \begin{tabular}{|c|c|c|c|c|c|c|c|}
        \hline
            & NC  & A & V & D & O & mean & model size (MB) \\
        \hline
        3DResnet\cite{hara2017learning}  & 0.560  & 0.866 & 0.259 & 0.052 & 0.929 & 0.533 & 325.22 \\ [0.025cm]
        \hline
        C3D\cite{tran2015learning}       & 0.972  & 0.965 & 0.920 & 0.895 & 0.989 & 0.948 & 33.56 \\ [0.025cm]
        \hline
        3DSE-SE           & 0.954  & 0.953 & 0.924 & 0.914 & 0.985 & 0.946 & 11.44 \\ [0.025cm]
        \hline
        3DSE              & 0.964  & 0.970 & 0.956 & 0.960 & 0.985 & \textcolor{red}{0.967} & 19.22 \\ [0.025cm]
        \hline
        3DSE+\acs{ACE}    & 0.988  & 0.992 & 0.963 & 0.951 & 0.993 & \textcolor{blue}{0.977} & 19.22 \\ [0.025cm]
        \hline
    \end{tabular} \vspace{-2.6mm}
\end{table}

\begin{figure}[t]
\centering
\includegraphics[width=1.00\textwidth]{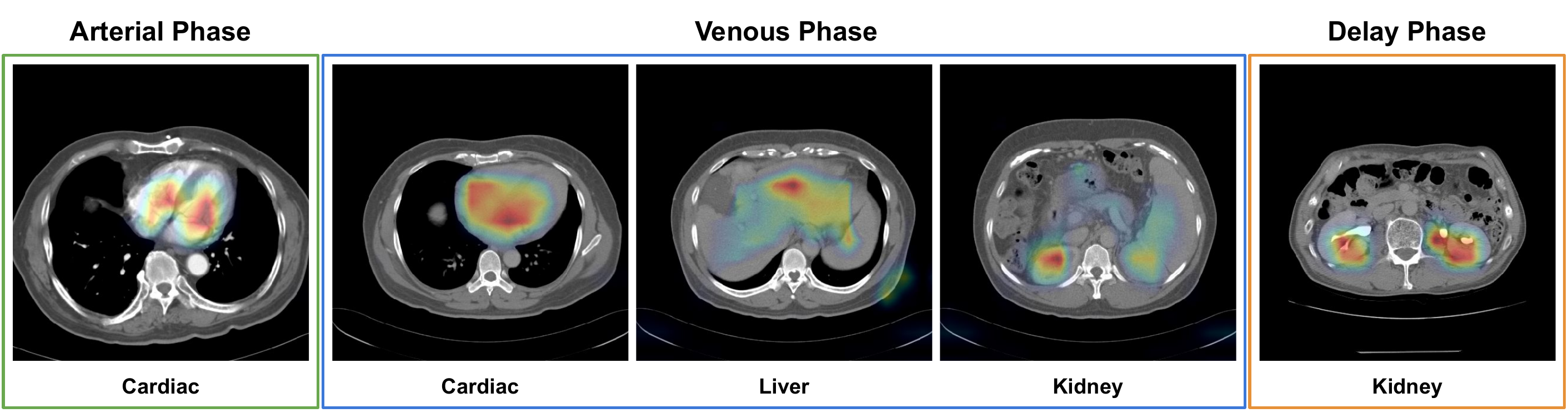}
\caption{Respond-CAM~\cite{zhao2018respond} visualizations of 3DSE from three different dynamic \acs{CT} scans. (\acs{A}) the 3DSE focuses on contrast accumulation in the cardiac region; (\acs{V}): 3DSE focuses on contrast remnants in the cardiac blood pool, liver portal veins, and kidney veins; (\acs{D}): 3DSE focuses on contrast accumulation in the ureters of the kidney.}
\label{fig:hm_vis}
\end{figure}

Shifting focus to across-model comparisons, \Tbl~\ref{tab:cnn_compare} compares our 3DSE model, with and without \ac{SE}, against other state-of-the-art 3D deep models \cite{hara2017learning,tran2015learning}. As can be seen, 3D-Resnet is nearly $17$ times larger than 3DSE and performs poorly, which we observed was due to overfitting. Moving down in model size, C3D~\cite{tran2015learning} performs better than 3D-Resnet, but is still unable to match 3DSE. If we remove the \ac{SE} layer from our 3DSE model, performance considerably suffers, which demonstrates that the \ac{SE} layer is important in achieving high performance. Despite this, performance still matches C3D even though a significantly smaller number of parameters are used. Finally, the last rows show 3DSE with and without the \ac{ACE} loss, with latter achieving the highest performance at a model size much smaller than competitors. Finally, as \Fig~\ref{fig:hm_vis} illustrates, the 3DSE model focuses on anatomical regions that are consistent with clinical practice. More visualizations can be found in our supplementary material. 

\begin{table}[!htb]
\centering
\caption{Study-level performance of text mining and 3DSE. Each row groups studies based on the number of dynamic \ac{CT} \acfp{SOI} they possess. Each column counts the number of studies based on how many scans were misclassified, if any. Best results for each \ac{SOI} number are marked in \textcolor{blue}{blue}.}
\label{tab:study_matrix}
    \begin{tabular}{|l||c|c|c||c|c|c||c|c|c|}
        \hline
           & \multicolumn{3}{c||}{Text Mining} & \multicolumn{3}{c||}{3DSE} & \multicolumn{3}{c|}{3DSE + ACE} \\
        \hline
           & 0 Errs.  & 1 Err. & $\geq 2$ Errs. & 0 Errs.  & 1 Err. & $\geq 2$ Errs. & 0 Errs.  & 1 Err. & $\geq 2$ Errs. \\ [0.025cm]
        \hline
        0 \acsp{SOI} & 35  & 8  & 10  & 47  & 4  & 2  & \textcolor{blue}{48}  & 5  & 0 \\ [0.025cm]
        \hline
        1 \ac{SOI} & 36  & 13 & 1   & 47  & 1  & 2  & \textcolor{blue}{49}  & 1  & 0 \\ [0.025cm]
        \hline
        2 \acsp{SOI} & 0   & 1  & 0   & 1   & 0  & 0  & \textcolor{blue}{1}   & 0  & 0 \\ [0.025cm]
        \hline
        3 \acsp{SOI} & 15  & 3  & 1   & 19  & 0  & 0  & \textcolor{blue}{19}  & 0  & 0 \\ [0.025cm]
        \hline
        4 \acsp{SOI} & \textcolor{blue}{101} & 6  & 1   & 95  & 12 & 1  & 97  & 10 & 1 \\ [0.025cm]
        \hline
        Total   & 186 & 32 & 13  & 209 & 16 & 6  & \textcolor{blue}{\textbf{214}} & 16 & 1 \\ [0.025cm]
        \hline
        Accuracy & 80.9\% & -- & -- & 90.5\% & -- & -- & 92.7\% & -- & -- \\ [0.025cm]
        \hline
    \end{tabular}
\end{table}

These boosts in scan-level performance are important, but arguably the study-level performance is even more important, as the ultimate goal is to identify and extract as many dynamic liver \ac{CT} studies as possible for downstream analysis. Thus, we also evaluate how many studies are correctly extracted, meaning all of their corresponding \acp{SOI} are correctly classified. As \Tbl~\ref{tab:study_matrix} demonstrates, $90.5\%$ of studies have all of their scans correctly classified by our 3DSE model. Including the wealky supervised data using the \ac{ACE} loss, we can further improve this to $92.7\%$. If we extrapolate these results to entire dataset of $7\,680$ studies, this means that the 3DSE model, armed with the \ac{ACE} loss, can correctly identify and extract $609$ more studies than the text mining approach. This is a significant boost of study numbers for any subsequent analyses.

\vspace{-0.2cm}
\section{Conclusion}
We presented a data curation tool to robustly extract multi-phase liver studies from a real-world and heterogenous hospital \ac{PACS}. This includes a streamlined, but powerful, 3DSE model and a principled \ac{ACE} loss designed to handle incompletely labelled data. Experiments demonstrated that our 3DSE model, along with the \ac{ACE} loss, can outperform both text mining and also more complex deep models. These results indicate that our vision-based approach can be an effective means to better curate large-scale clinical datasets. Future work includes evaluating our approach in other clinical scenarios, as well as investigating how to harmonize text-mined features with our visual-based system. 

\vspace{-0.2cm}
\bibliographystyle{splncs}
\bibliography{bibliography}
\end{document}


\title{CT Data Curation for Liver Patients: Phase Recognition in Dynamic Contrast-Enhanced CT -Supplementary Material}

\author{Bo Zhou\inst{1,2} \and Adam P. Harrison\inst{2} \and Jiawen Yao\inst{2} \and Chi-Tung Cheng\inst{4} \and Jing Xiao\inst{3} \and Chien-Hung Liao\inst{4} \and Le Lu\inst{2}}
\institute{Biomedical Engineering, Yale University, New Haven, CT, USA \and PAII Inc., Bethesda, MD, USA \and PingAn Technology, Shenzhen, China \and Chang Gung Memorial Hospital, Linkou, Taiwan, ROC}
 
\authorrunning{B. Zhou, A. P. Harrison, etc} 
\titlerunning{Data Curation for Patients: Phase Recognition} 

\maketitle              




\section{Additional Results}

\begin{figure}[!htb]
\centering
\includegraphics[width=0.95\textwidth]{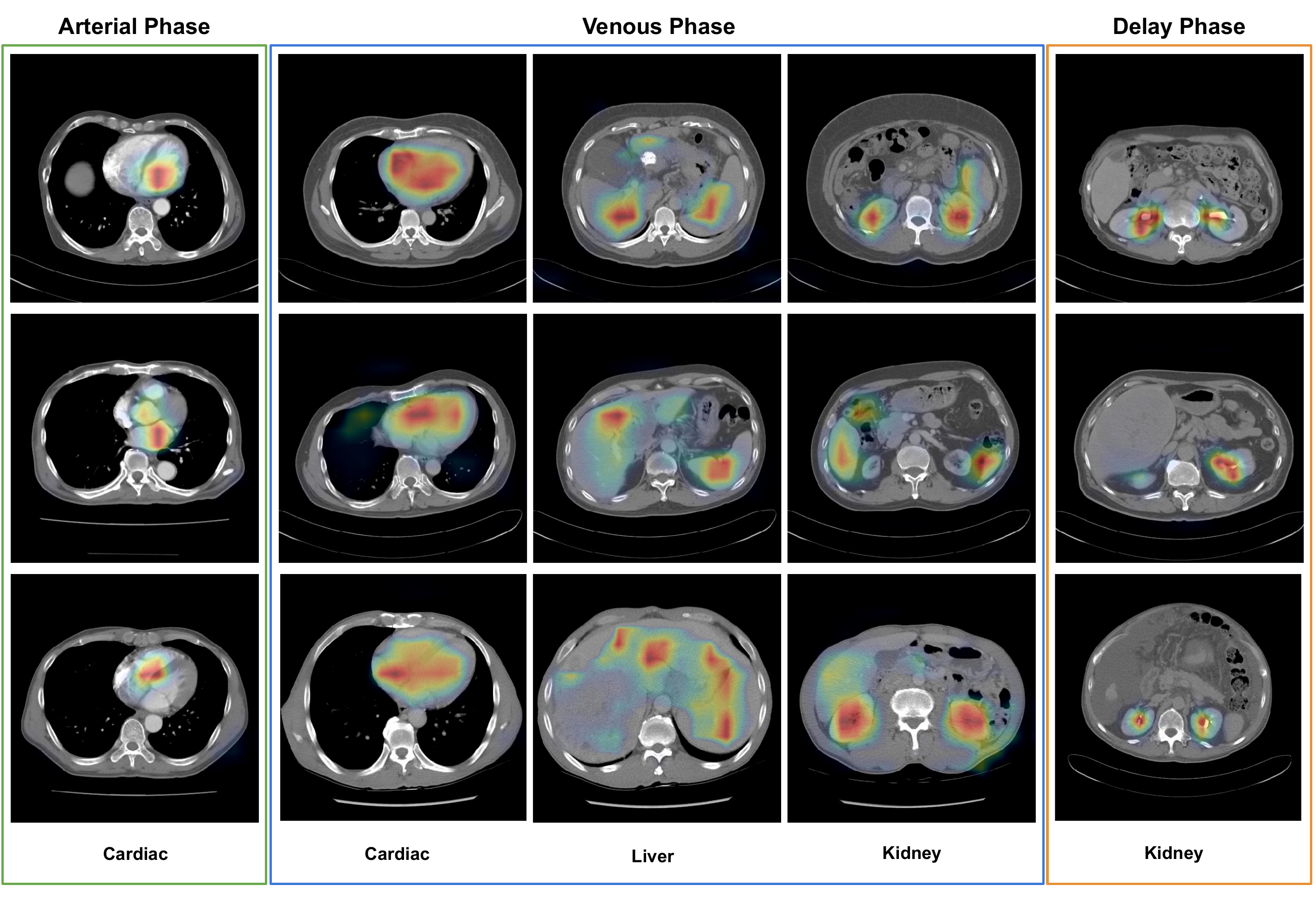}
\caption{Additional respond-CAM visualizations of 3DSE.}
\label{fig:hm_vis_add}
\end{figure}

\newpage
\section{Text Mining Rules}

\newcommand{\tabincell}[2]{\begin{tabular}{@{}#1@{}}#2\end{tabular}} 
\begin{table}[htb!]
\scriptsize
\centering
\caption{The text-mining rules for generating the initial annotation on our dynamic CT dataset. The study description, series description, and protocol DICOM tags are used for mining. Please note the study description and protocol are shared across all series in a study.}
\label{tab:scan_confusion_matrix}
    \begin{tabular}{|c|c|c|}
        \hline
        \multicolumn{2}{|c|}{\textbf{Class}} & \textbf{Rules}  \\ [0.025cm]
        \hline
        \multicolumn{2}{|c|}{\textbf{arterial phase}} & \tabincell{c}{``arterial''/``liver A''/``Liver -A''/``Artery''\\ /``HAP''/``Liver 3P A''/``Liver 3P C+ A''\\ /``A Phase''/``A-Phase''/``Liver 3P C-+ A.''/``CTA''\\ /``Liver 4P A''/``30s'' in series description} \\ [0.035cm]
        \hline \hline
        \multicolumn{2}{|c|}{\textbf{venous phase}} & \tabincell{c}{``venous''/``Liver 3P V''/``Liver 3P C+ V''/``Liver 3P +H''\\ /``LIVER H''/``Liver -V''/``Liver V''/``Portal''\\ /``Liver -V''/``Liver P.''/``Vena''\\ /``PVP''/``Phase H.''/``H./Phase''\\ /``V phase''/``phase V''/``V-phase''\\ /``Liver 3P C-+ V.''/``CTV''\\ /``Liver 4P V''/``70s'' in series description} \\ [0.035cm]
        \hline \hline
        \multicolumn{2}{|c|}{\textbf{delay phase}} & \tabincell{c}{``Liver 3P D''/``Liver D.''/``Liver -D''\\/``delay''/``Liver 3P C-+ D.''/``Liver 3P C+ D''\\ /``D-phase''/``Liver 4P D''/``DP''/``180s''\\ /``EQP'' in series description} \\ [0.035cm]
        \hline \hline
        \multicolumn{2}{|c|}{\textbf{non-contrast phase}} & \tabincell{c}{``C-''/``PRECONTRAST''/``Abd-pelvis without contrast''\\ /``Non:Contrast''/``Non Contrast''/``Non-Contrast''\\/``NoC'' in series description} \\ [0.035cm]
        \hline \hline
        \multicolumn{2}{|c|}{\textbf{contrast}} & \tabincell{c}{``ABD C+''/``A C+''/``abdomen C+''\\ /``AbdPel C''/``Body C+''/``with contrast''\\ /``POSTCONTRAST'' in series description} \\ [0.035cm]
        \hline \hline
        \multirow{14}{*}{\textbf{other}} & \textit{CTAP} & \tabincell{c}{``CTAP'' in protocol/series description} \\ [0.035cm] 
        \cline{2-3}
         & \textit{guide} & \tabincell{c}{``guide''/``BX''/``POST''\\ in study description / series description} \\ [0.035cm] 
        \cline{2-3}
         & \textit{scano} & \tabincell{c}{``Topo''/``scano''/``scout''/``surview'' in series description} \\ [0.035cm]
        \cline{2-3}
         & \textit{MIP} & \tabincell{c}{``MIP'' in series description} \\ [0.035cm]
        \cline{2-3}
         & \textit{volume} & \tabincell{c}{``volume'' in series description} \\ [0.035cm]
        \cline{2-3}
         & \textit{monitor} & \tabincell{c}{``monitor'' in series description} \\ [0.035cm]
        \cline{2-3}
         & \textit{coronal} & \tabincell{c}{``cor'' in series description} \\ [0.035cm]
        \cline{2-3}
         & \textit{oblique} & \tabincell{c}{``obl'' in series description} \\ [0.035cm]
        \cline{2-3}
         & \textit{reformatted} & \tabincell{c}{``reformatted'' in series description} \\ [0.035cm]
        \cline{2-3}
         & \textit{brain} & \tabincell{c}{``brain'' in series description} \\ [0.035cm]
        \cline{2-3}
         & \textit{chest} & \tabincell{c}{``lung''/``CXR''/``chest'' in series description} \\ [0.035cm]
        \cline{2-3}
         & \textit{pelvic} & \tabincell{c}{``pelvic''/``Pelvis'' in series description} \\ [0.035cm]
        \cline{2-3}
         & \textit{3 phases in 1} & \tabincell{c}{``3 Phase Liver''/``120CC/3CC/SEC''\\ /``4cc sec''/``Tri-Phase Liver'' in series description} \\ [0.035cm]
        \hline
    \end{tabular}
\end{table}
